\documentclass[11pt]{article}
\usepackage{graphicx}
\usepackage{epsfig}

\newcommand{\BABARPubYear}    {00}

\newcommand{\BABARConfNumber} {04}
\newcommand{\SLACPubNumber} {8526}

\usepackage{relsize}
\def\babar{\mbox{\slshape B\kern-0.1em{\smaller A}\kern-0.1em
    B\kern-0.1em{\smaller A\kern-0.2em R}}}

\def\epem       {\ensuremath{e^+e^-}}

\def\mumu       {\ensuremath{\mu^+\mu^-}}

\def\pipi  {\ensuremath{\pi^+\pi^-}}

\def\Kbar  {\kern 0.2em\overline{\kern -0.2em K}{}}

\def\Kzb   {\ensuremath{\Kbar^0}}
\def\KzKzb {\ensuremath{K^0 \kern -0.16em \Kzb}}

\def\Dbar  {\kern 0.2em\overline{\kern -0.2em D}{}}

\def\Dzb   {\ensuremath{\Dbar^0}}
\def\DzDzb {\ensuremath{D^0 {\kern -0.16em \Dzb}}}

\def\Bbar  {\kern 0.18em\overline{\kern -0.18em B}{}}

\def\Bzb   {\ensuremath{\Bbar^0}}

\def\BB    {\ensuremath{B\Bbar}} 
\def\BzBzb {\ensuremath{B^0 {\kern -0.16em \Bzb}}}

\def\jpsi  {\ensuremath{{J\mskip -3mu/\mskip -2mu\psi\mskip 2mu}}} 
\def\psitwos {\ensuremath{\psi{(2S)}}}
\mathchardef\Upsilon="7107
\def\Y#1S{\ensuremath{\Upsilon{(#1S)}}}

\def\FourS {\Y4S}

\mathchardef\Deltares="7101
\mathchardef\Xi="7104
\mathchardef\Lambda="7103
\mathchardef\Sigma="7106
\mathchardef\Omega="710A
\def\Deltabar   {\kern 0.25em\overline{\kern -0.25em \Deltares}{}}
\def\Lbar {\kern 0.2em\overline{\kern -0.2em\Lambda\kern 0.05em}\kern-0.05em{}}
\def\Sigbar{\kern 0.2em\overline{\kern -0.2em \Sigma}{}}
\def\Xibar{\kern 0.2em\overline{\kern -0.2em \Xi}{}}
\def\Obar{\kern 0.2em\overline{\kern -0.2em \Omega}{}}
\def\Nbar{\kern 0.2em\overline{\kern -0.2em N}{}}
\def\Xbar{\kern 0.2em\overline{\kern -0.2em X}{}}

\def\BR{{\ensuremath{\cal B}}}

\def\ev   {\ensuremath{\rm \,e\kern -0.08em V}}
\def\kev  {\ensuremath{\rm \,ke\kern -0.08em V}} 
\def\mev  {\ensuremath{\rm \,Me\kern -0.08em V}} 
\def\gev  {\ensuremath{\rm \,Ge\kern -0.08em V}} 
\def\gevc {\ensuremath{{\rm \,Ge\kern -0.08em V\!/}c}} 
\def\tev  {\ensuremath{\rm \,Te\kern -0.08em V}}
\def\mevc {\ensuremath{{\rm \,Me\kern -0.08em V\!/}c}} 
\def\gevcc{\ensuremath{{\rm \,Ge\kern -0.08em V\!/}c^2}} 
\def\mevcc{\ensuremath{{\rm \,Me\kern -0.08em V\!/}c^2}} 

\def\in   {\ensuremath{\rm \,in}}

\def\cm   {\ensuremath{\rm \,cm}}

\def\mm   {\ensuremath{\rm \,mm}}

\def\mus  {\ensuremath{\rm \,\mus}}

\def\mus        {\ensuremath{\,\mu{\rm s}}}    
\def\gsim{{~\raise.15em\hbox{$>$}\kern-.85em
          \lower.35em\hbox{$\sim$}~}}
\def\lsim{{~\raise.15em\hbox{$<$}\kern-.85em
          \lower.35em\hbox{$\sim$}~}}

\def\CP                 {\ensuremath{C\!P}}

\def\pep2{PEP-II}

\def\chic#1{\ensuremath{\chi_{c#1}}} 

\newcommand{\dedx}{\ensuremath{\mathrm{d}\hspace{-0.1em}E/\mathrm{d}x}}

\newcommand{\eqref}[1]{Eq.~(\ref{eq:#1})}

\newcommand{\epjc}      [1]  {{Eur.\ Phys.\ Jour.\ C~{\bf #1}}}

\newcommand{\prl}       [1]  {{Phys.\ Rev.\ Lett.\ {\bf #1}}} 

\def\jetset74   {\mbox{\tt Jetset \hspace{-0.5em}7.\hspace{-0.2em}4}}

\def\psitwos {\ensuremath{{\psi(2S)}}}
\def\chic {\ensuremath{{\chi_c}}}
\def\chicone {\ensuremath{{\chi_{c1}}}}
\def\chictwo {\ensuremath{{\chi_{c2}}}}

\long\def\inst#1{\par\nobreak\kern 4pt\nobreak
    {\it #1}\par\vskip 10pt plus 3pt minus 3pt}

\begin{document}
{\pagestyle{empty}

\begin{flushright}
\babar-CONF-\BABARPubYear/\BABARConfNumber \\
SLAC-PUB-\SLACPubNumber
\end{flushright}


\par\vskip 3cm

\begin{center}
\Large \bf Measurement of inclusive production of charmonium states
in \boldmath $B$ meson decays
\end{center}
\bigskip

\begin{center}
\large The \babar\ Collaboration\\
\mbox{ }\\
July 25, 2000
\end{center}
\bigskip \bigskip

\begin{center}
\large \bf Abstract
\end{center}
We reconstruct the charmonium mesons \jpsi, \psitwos\ and \chic\ using
a sample of $8.46\times 10^6$
\BB\ events collected by the \babar\ detector operating at
$e^+ e^-$ center of mass energies near the \FourS\ resonance.
By measuring rates relative to the branching fraction
of the \jpsi,
we obtain preliminary inclusive $B$ branching fractions 
of $(0.25\pm0.02 \pm0.02)$\% to the \psitwos\ and 
$(0.39\pm0.04 \pm0.04)$\% to the
\chicone, and set a 90\% confidence level limit of 0.24\% on
decays through the \chictwo.

\vfill
\begin{center}
Submitted to the XXX$^{th}$ International 
Conference on High Energy Physics, Osaka, Japan.
\end{center}

\newpage
}

\begin{center}
\small

The \babar\ Collaboration
\bigskip

B.~Aubert,
A.~Boucham,
D.~Boutigny,
I.~De Bonis,
J.~Favier,
J.-M.~Gaillard,
F.~Galeazzi,
A.~Jeremie,
Y.~Karyotakis,
J.~P.~Lees,
P.~Robbe,
V.~Tisserand,
K.~Zachariadou
\inst{Lab de Phys.\ des Particules, F-74941 Annecy-le-Vieux, CEDEX, France}
% ...

\end{center}\newpage

\setcounter{footnote}{0}

\section{Introduction}
\label{sec:Introduction}

Observation of the charmonium mesons \jpsi, \psitwos\ and \chic\ is a critical
component of the measurement of \CP\ violation in $B$ decays at \babar\ and other
facilities.  Measurements of the inclusive branching fractions
of $B$ mesons to these states are made using the decays
$\jpsi\rightarrow\ell^+\ell^-$, $\psitwos\rightarrow\ell^+\ell^-$, 
$\psitwos\rightarrow\pipi\jpsi$
and $\chic \rightarrow \gamma\jpsi$, 
with $\jpsi\rightarrow\ell^+\ell^-$ in the latter two cases;
the leptons $\ell$ may be either electrons or muons.
The production rates of the \psitwos\ and \chic\ mesons are measured relative
to the \jpsi\ in order to eliminate common systematic errors in tracking and
particle identification.

\section{The \babar\ detector and dataset}
\label{sec:babar}
The data used in this analysis were collected with the \babar\ detector
at the \pep2\ storage ring, which collides 9.0\gev\ electrons
with 3.1\gev\ positrons.   
A luminosity of 7.7~$fb^{-1}$ was collected on the \FourS\ 
resonance, $\sqrt s = 10.58$\gev\ (``on-resonance''), 
with an additional 1.2~$fb^{-1}$ collected at  $\sqrt s = 10.54$\gev,
below the threshold for \BB\ production (``off-resonance'').
The off-resonance data set
is used to statistically subtract the non-\BB\ component of signals.

The \babar\ detector is described elsewhere~\cite{ref:babar}.
The tracking system, which is used for pattern recognition and
reconstruction of the momenta of charged particles, consists of
two sub-detectors.
The inner 
is a five-layer double sided
silicon vertex tracker (SVT)
which gives precision spatial information for
all charged particles and is the primary detection device for low momentum
charged particles. It is surrounded by a 40-layer drift chamber (DCH),
which 
provides measurements of track momenta. 

The primary sources of information used in 
lepton identification are the calorimeter and the muon system.
The DCH also provides \dedx\ information which is
used in the identification of 
electrons.  
The calorimeter is constructed from CsI crystals and, in addition to 
distinguishing electrons, hadrons and muons on the basis of their
energy depositions, is used to locate and measure photons.  The muon
system, known as the Instrumented Flux Return (IFR) in \babar, consists of 
resistive plate chambers interleaved with iron plates.  There are 19
layers in the central region of the detector, 18\in\ the endcap regions.


\section{Event selection}
\label{sec:eventsel}

The event selection criteria are designed to select \BB\ decays with high
efficiency while suppressing lepton pairs, two-photon events, and 
interactions with the beam pipe or residual gas in the beam line.  
We require the event to have:
\begin{itemize}
\item at least four charged tracks in the active volume
of the tracking system; 
\item total energy (charged tracks plus photons) greater than
5\gev;
\item a primary vertex within 5\mm\ of the beam interaction region; and
\item a normalized second-order Fox-Wolfram moment~\cite{ref:fox} 
less than 0.7.
\end{itemize}


\noindent
A total of $(8.46 \pm 0.14)\times 10^6$ \BB\ events satisfy these 
criteria.  This value is obtained by scaling the number of events
selected in the off resonance sample by the ratio of muon pairs
(luminosities) and subtracting it from the total number of events
selected in the on-resonance sample.  The uncertainty is dominated
by variations in the ratio of selected events to muon pairs in the 
off-resonance running periods throughout the data collection period.


\section{Reconstruction of charmonium mesons}
\label{sec:reco}

\subsection{Reconstruction of \boldmath \jpsi}

\jpsi\ candidates are formed from pairs of charged tracks in the active
fiducial volume of the tracking system and calorimeter. Tracks are
required to pass within 1.5\cm\ of the beam line and to include
information from the DCH.


Both tracks are required to satify lepton-identification criteria.
Electrons must have an energy deposition in the calorimeter 
of at least 75\% of that expected from the track's momentum and a
shape consistent with an electromagnetic shower.
Muons must have an energy deposition in the calorimeter of less than 
0.5\gev---approximately twice the expectation for a minimum ionizing
track---and traverse an amount of material that is within two
interaction lengths of 
that expected for a muon of the measured momentum.  The pattern of 
energy deposition in the IFR must be narrow, perpendicular to the flight
path of the track, and the location of the hits must have an acceptable
$\chi^2$ when fit to the trajectory extrapolated from the DCH.

The two tracks are fit to a single vertex, if possible, before
calculating the mass of the candidate \jpsi. Otherwise, 
the mass is derived from a simple sum of their four-vectors.

\begin{figure}
\epsfig{figure=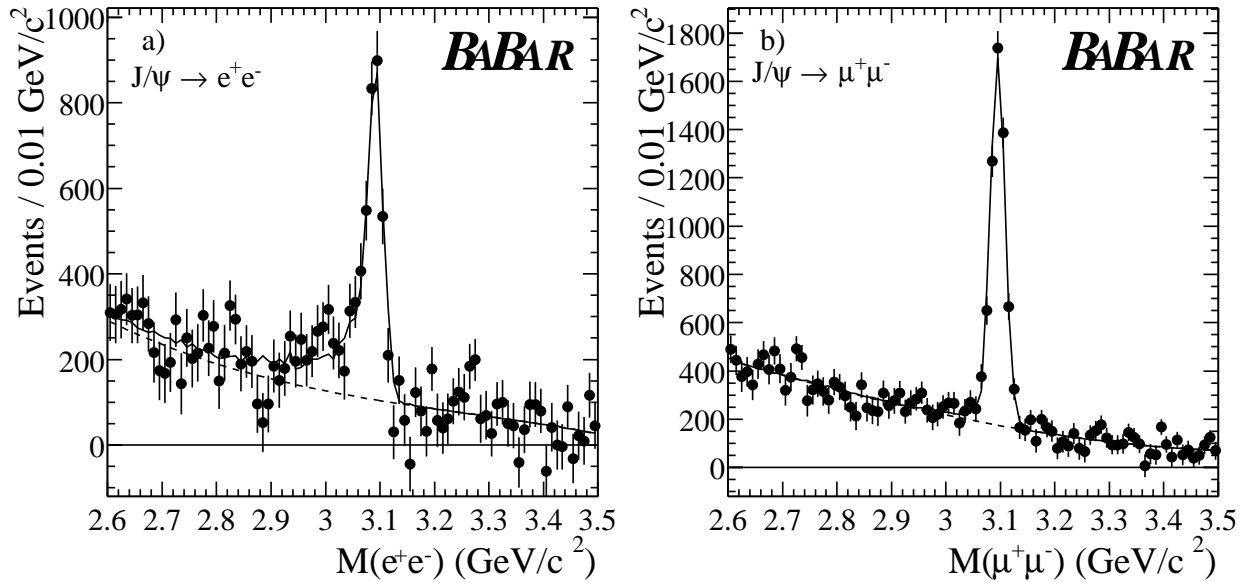,width=\linewidth}
\caption{Mass distribution of \jpsi\ candidates to (a) electron pairs
and (b) muon pairs after the continuum subtraction.}
\label{fig:jpsimass}
\end{figure}

Histograms of mass distributions are accumulated for both the on-
and off-resonance data sets. These are then subtracted, after weighting for
the difference  in luminosities, to give ``continuum subtracted''
distributions.
Figure~\ref{fig:jpsimass} shows
the resulting mass distributions for $\jpsi\rightarrow\epem$ and
$\jpsi\rightarrow\mumu$ candidates.

The number of \jpsi\ mesons is extracted from the mass distribution
using a probability distribution function (p.d.f.) derived from a simulation
that includes final state radiation
and bremsstrahlung.  The mass resolution of the detector
simulation is increased from 11.0 to 13.3\mevcc\ to match
the observed value.

A systematic error on the fit to the \epem\ final state is obtained by
varying by $\pm 10$\% the number of \jpsi\ mesons 
in which an electron undergoes bremsstrahlung in the simulation.
This variation, which reflects possible differences between the
detector simulation and reality, translates to a 3.5\% systematic error.

Systematic errors for both final states are obtained by modifying the 
selection criteria in two ways.  First, the
momentum of the \jpsi\ candidate
in the \FourS\ center of mass (CM) system, $p^*$,
is required
to be less than 2.0\gevc; all \jpsi\ mesons from $B$ decays
should pass.  Second, the number of charged tracks required is increased
to five in order to reject remaining lepton pair events.  These studies
produce systematic errors of 1.3\% for \epem\ and 1.6\% for \mumu.

We find $4920 \pm 100 \pm 180$ $\jpsi\rightarrow\epem$ and
$5490 \pm 90 \pm 90$ $\jpsi\rightarrow\mumu$ events with mass
greater than 2.6\gevcc.


\subsection{Reconstruction of \boldmath \psitwos}

The reconstruction of the \psitwos\ through its decay to \epem\ or
\mumu\ is similar to that of the \jpsi.  In order to reduce the
relatively higher background rates, additional selection criteria
are applied to the \psitwos\ candidate.  First, its momentum in the
CM system ($p^*$) must be less than 1.6\gevc, true for all \psitwos\
mesons from $B$ decays.  Second, the two tracks must be consistent with
coming from a common vertex with a probability greater than 1\%.
Finally, more stringent particle identification criteria are applied. 
Muon candidates must penetrate to within one interaction length of 
the expected amount of material and at least one electron candidate
must have an energy deposition between 0.88 and 1.30 times its momentum.

\begin{figure}
\epsfig{figure=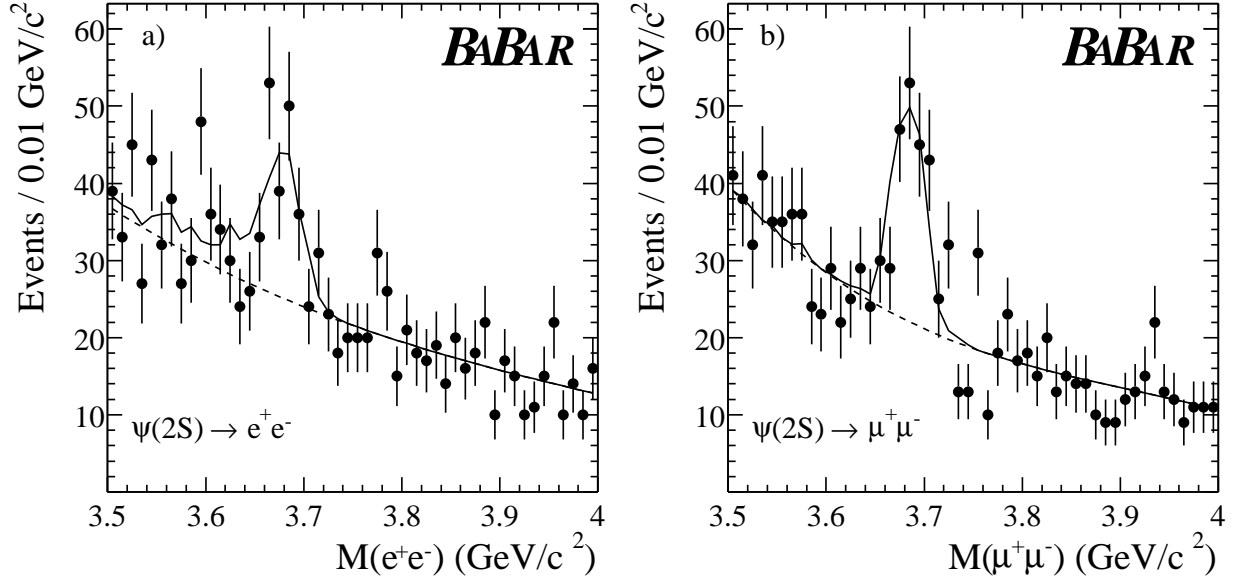,width=\linewidth}
\caption{Mass distribution of \psitwos\ candidates to (a) electron pairs
and (b) muon pairs.}
\label{fig:psi2sll}
\end{figure}

The number of signal events is extracted from the mass distribution  
using a p.d.f. from simulation, as for the \jpsi.  We do not perform a 
continuum subtraction in this case, as the requirement on $p^*$ is expected
to remove most backgrounds from the continuum.  Variation of the 
amount of bremsstrahlung in the detector simulation leads to a 1.7\%
systematic error on the number of $\psitwos\rightarrow\epem$ events,
although this error is  negligible compared to the statistical errors
in this case.
We find $131 \pm 29 \pm 2$ $\psitwos\rightarrow\epem$ and
$125 \pm 19$ $\psitwos\rightarrow\mumu$ events 
(Fig.~\ref{fig:psi2sll}).

Due to their lower momentum, the pion candidates used to reconstruct
$\psitwos\rightarrow \pipi \jpsi$ are not required to 
have DCH information. 
Pairs of pions with mass between 0.45 and 0.59\gevcc\ 
are
combined with \jpsi\ candidates 
with mass between 2.60 and 3.13\gevcc\ for \epem\ and 
3.06 and 3.13\gevcc\ ($-2.8$ to $+2.5$ standard deviations) for \mumu.
The wider window for \epem\ is necessitated by the larger radiative tail
in this final state.
The above requirements on $p^*$ and on vertex quality are applied in this
final state as well.

\begin{figure}
\epsfig{figure=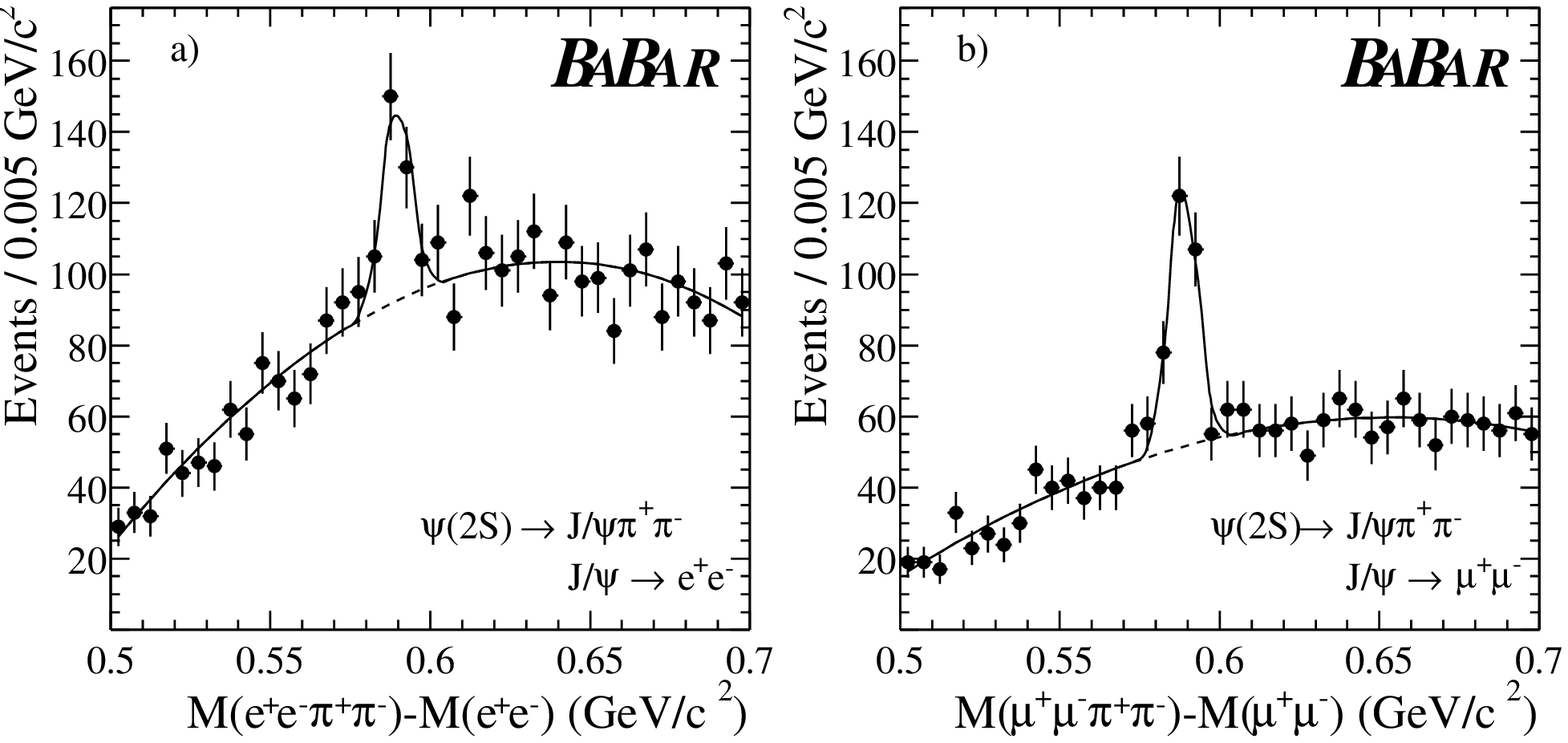,width=\linewidth}
\caption{Mass difference between the \psitwos\ and \jpsi\ candidates, 
where $\psitwos\rightarrow \pipi \jpsi$ and (a) $\jpsi \rightarrow \epem$
and (b) $\jpsi \rightarrow \mumu$.}
\label{fig:psi2spipi}
\end{figure}

To reduce the impact of the radiative tail and the mass resolution of the
\jpsi\ candidate, the number of \psitwos\ events is extracted from a fit to
the distribution of
the mass difference between the \psitwos\ and the \jpsi\ 
(Fig.~\ref{fig:psi2spipi}), with a Gaussian distribution for the signal
and a Chebychev polynomial for background.
In candidates with 
$\jpsi\rightarrow\epem$, we obtain $126 \pm 44$ \psitwos\ events, 
while for $\jpsi\rightarrow\mumu$, we obtain $162 \pm 23$.

\subsection{Reconstruction of \boldmath \chic}

\chic\ mesons are reconstructed through the radiative decay 
$\chic\rightarrow\gamma\jpsi$.  The selected photon candidates must be
reconstructed within the same fiducial region used for leptons and have
an energy between 0.20 and 0.65\gev\ in the CM system.
The shape of the energy deposition must be consistent with an
electromagnetic shower.  To reject photons from nearby hadronic showers, 
the photon must be at least $9^\circ$ from the nearest charged track
in the calorimeter.
The substantial combinatorial background from $\pi^0$ decays is reduced
by rejecting any photon that, when combined with any other photon,  
forms a $\pi^0$ candidate with mass between 0.109 and 0.147\gevcc\
($-4.0$ to $+2.0$ standard deviations about the $\pi^0$ mass).


In selecting the \jpsi, muon candidates must satisfy the more stringent
requirement that they penetrate to within one interaction length of 
the expected amount of material.
\jpsi\ candidates with mass between 3.05 and 3.14\gevcc\ are combined
with photon candidates to form \chic\ candidates.  

\begin{figure}
\epsfig{figure=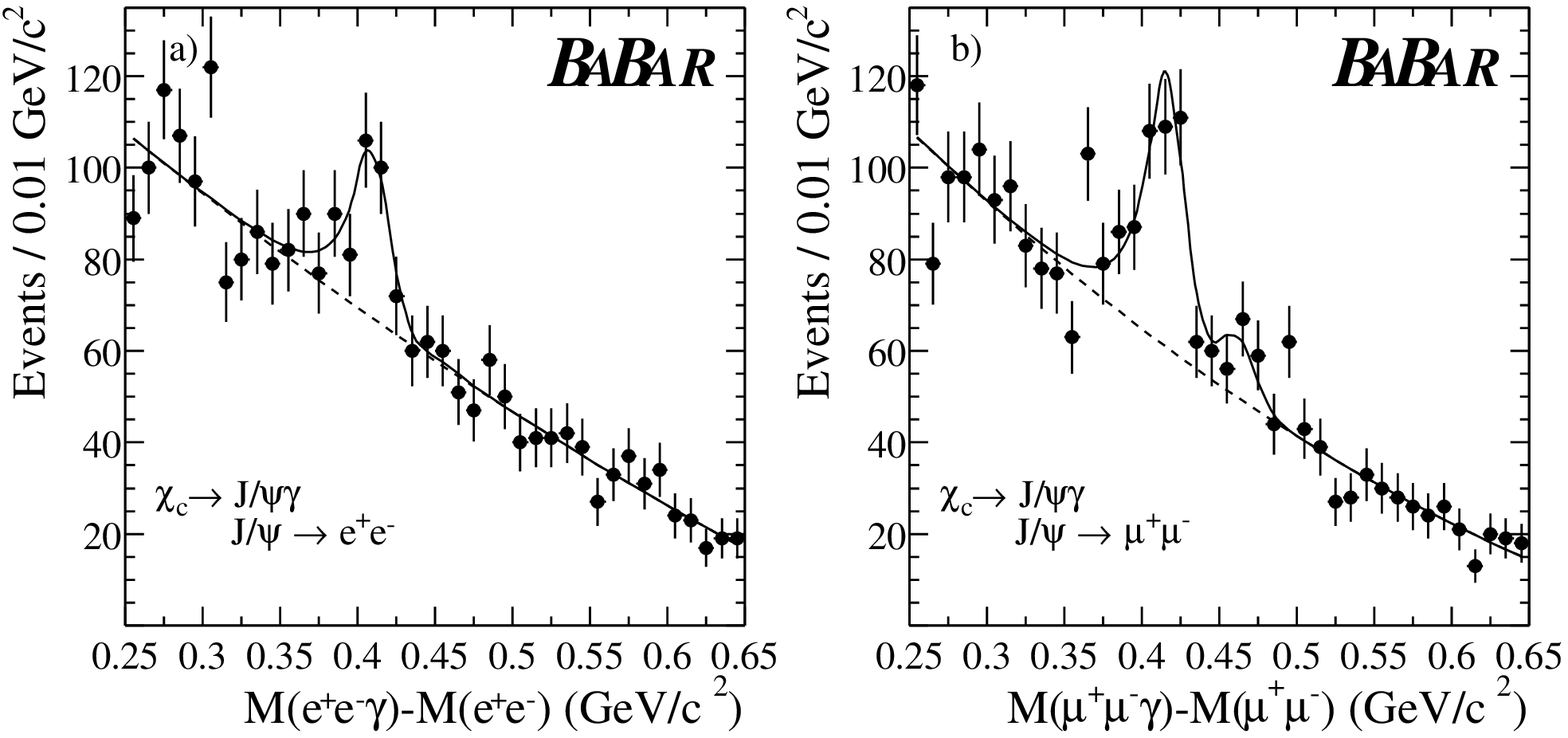,width=\linewidth}
\caption{Mass difference between the \chic\ and \jpsi\ candidates, 
where $\chic\rightarrow \gamma \jpsi$ and (a) $\jpsi \rightarrow \epem$
and (b) $\jpsi \rightarrow \mumu$.}
\label{fig:chic}
\end{figure}

As for the \psitwos, the number of events is extracted by fitting
the distribution of the mass difference between the \chic\ and \jpsi\
candidates (Fig.~\ref{fig:chic}).  The resolution is dominated by the
angular and energy resolutions for the photon, and includes a low-side
tail due to preshowering before the calorimeter. 
\chic\ signals are extracted from a fit
with a ``Crystal Ball'' function, which includes this feature.
The parameter that describes the tail is fixed to the value determined
from a complete detector simulation\cite{ref:cb}.


A 4.2\% systematic error due to a possible mismatch between the
fitting function and the true distribution is estimated from fits to 
simulated data.  An additional systematic error 
is evaluated by requiring $p^* < 2.0$\gevc\ for
the \jpsi\ candidate.  The resulting variation in
the signal,  9.3\% for \epem\ and
2.0\% for \mumu, although not inconsistent with a statistical
fluctuation, is taken as a systematic error.

We simultaneously fit for a \chicone\ and a possible \chictwo\ component.  The only
additional free parameter is the number of \chictwo\ candidates:
the mass
difference between the \chictwo\ and the \chicone\ is fixed to the
Particle Data Group value and the resolution and tail parameters are
fixed to those of the \chicone\cite{ref:pdg}.


We find $129 \pm 26 \pm 13$ \chicone\ and $3 \pm 21$ \chictwo\ 
events in candidates in which $\jpsi\rightarrow\epem$ and 
 $204 \pm 47 \pm 12$ \chicone\ and $47 \pm 21$ \chictwo\ 
in candidates in which $\jpsi\rightarrow\mumu$.

\section{Calculation of branching fractions}
\label{sec:branchrat}

\subsection{\boldmath $B \rightarrow \jpsi X$}
\label{subsec:jpsi}

The number of \jpsi\ mesons produced in our data sample is related to the
number extracted from the fit by a correction for the reconstruction
efficiency and by the branching fraction for $\jpsi\rightarrow\ell^+\ell^-$.
We average the published $\jpsi\rightarrow\epem$ and 
$\jpsi\rightarrow\mumu$ branching fractions, assuming a common
systematic error, to get $(5.91\pm 0.10)$\%\cite{ref:pdg}.

A complete detector simulation is used to extract the probability
that both leptons are reconstructed as tracks of sufficient quality
within the fiducial volume, with a combined mass greater than 2.6\gevcc\
(the region used in the fit to the data).  

The accuracy of the detector simulation in reproducing 
tracking efficiency has been studied
in $D^0 \rightarrow K^-\pi^+\pipi$ decays and in $\tau$ pairs in which
one $\tau$ decays to a single charged track and the other decays to
three charged track~\cite{ref:babar}.  From these studies, a uncorrelated
systematic error of 
2.5\% per track is assigned to the overall efficiency.

The probability of the
leptons being identified correctly is determined 
using ``control samples''---sets of tracks in which
the nature of the particle is known without the use of the particle identification
systems.  For both electrons and muons, we use 
lepton pairs and radiative lepton pairs produced in two-photon
events.
A track is  included in 
the control sample if the kinematics of the event are consistent with
expectations and if the other track in the event passes strict particle
identification requirements.
The control sample tracks are then used to evaluate
the efficiency of the selection criteria as a function of momentum and
location in the detector.
A value for the branching fraction $B\rightarrow\jpsi\ X$ will be determined once this
study has been completed.

\subsection{\boldmath $B \rightarrow \psitwos X$}
\label{subsec:psi2s}

To minimize systematic errors from tracking and lepton identification, we
calculate the ratio of \psitwos\ and \jpsi\ inclusive branching fractions using
the ratio of observed \psitwos\ to \jpsi\ mesons. 
For the purposes of this calculation, we refit the \jpsi\ signal after applying
the more stringent particle identification criteria used in reconstructing
the \psitwos.  
Simulation studies indicate that the probability for the  
$\psitwos\rightarrow\ell^+\ell^-$ leptons to be reconstructed and
pass the identification criteria is equal to that for the
$\jpsi\rightarrow\ell^+\ell^-$ decay.

The ratio of \psitwos\ to \jpsi\ inclusive branching fractions is equal to the
ratio of observed mesons, corrected for acceptance, the branching fractions 
$\jpsi \rightarrow  \ell^+\ell^-$ and $\psitwos \rightarrow  \ell^+\ell^-$
and a correction for the efficiency of the vertexing requirement.  
The vertexing efficiency is found by simulation to be 0.941 with a 2\%
systematic error, which is determined from vertexing studies of the \jpsi.

For the branching fraction of $\psitwos\rightarrow\ell^+\ell^-$ we 
average the two Particle Data Group values and obtain
$(0.90 \pm 0.12)$\%.  This is the dominant systematic error.

The ratio of \psitwos\ to \jpsi\ inclusive branching fractions is calculated
from the $\pipi \jpsi$ final state data
in a similar fashion.  It is equal to the
number of reconstructed \psitwos\ mesons divided by the number of \jpsi\
mesons reconstructed within the tighter mass windows used in this analysis,
with corrections for the pion reconstruction probability, the vertexing
efficiency and the $\psitwos\rightarrow \pipi\jpsi$ branching fraction. The
probability for the two pions to be reconstructed within the fiducial
requirements and
the  mass window is 0.524.  The efficiency of the requirement for a good
vertex is found by simulation to be 0.89 with a 4\% systematic error. The
uncertainty in the published branching fraction for $\psitwos\rightarrow \pipi\jpsi$, 
$(31.0 \pm 2.8)\%$, is the largest
component of the systematic error.

The two decay modes give consistent values for the ratio of
inclusive branching fractions
$\BR(B\rightarrow\psitwos X)/\BR(B\rightarrow\jpsi X)$
and are combined to give an
overall value in
Table~\ref{tab:brsum}.  The ratio is multiplied by the Particle Data Group
value for $\BR(B\rightarrow\jpsi X)$, $(1.15 \pm 0.06)$\%, to obtain the
inclusive \psitwos\ branching fraction.

\subsection{\boldmath $B \rightarrow \chic X$}
\label{subsec:chic}

The ratio of \chicone\ to \jpsi\ inclusive branching fractions 
is computed from the number of reconstructed \psitwos\ mesons divided by the
number of \jpsi\ mesons reconstructed within the tighter mass windows 
and tighter particle identification criteria
used in
this analysis, 
the published $\chicone\rightarrow \gamma\jpsi$
branching fraction,
and a correction for the photon reconstruction probability.

The efficiency for the photon reconstruction is calculated
from the simulation
to be 0.550. A systematic
error of 1.6\% is obtained by varying the 
calorimeter 
simulation resolution and
energy scales over bounds determined by studies of $\pi^0$
reconstruction. 
The branching
fraction for $\chicone\rightarrow\gamma\jpsi$, $(27.3\pm1.6)$\%, 
contributes a 5.8\% systematic error.  The systematic errors on the
fit have been described earlier.

To obtain a limit on the production of the \chictwo, the fit results from 
Fig.~\ref{fig:chic} are  recast  
in terms of the ratio of \chictwo\ to \jpsi\ inclusive branching fractions.  
The likelihood function for each fit is then normalized to unity for
$N_{\chictwo}\ge 0$ and multiplied together to give the combined 
likelihood.  The 90\% confidence level is taken
to be the ratio of branching fractions that bounds 90\% of the area
of the likelihood function. 

The statistical errors from the fit lead to a 90\% 
confidence limit on  $\BR(B\rightarrow\chictwo)/\BR(B\rightarrow\jpsi)$ of
0.186.  However,
there is an 8\% systematic error in converting the fits into limits 
on \chictwo\ production due to the uncertainty in the 
published branching fraction for $\chictwo\rightarrow\gamma\jpsi$.  We include this
uncertainty by increasing the limit by 1.28 times this
error, giving a limit of 0.205.

The results for \chicone\ and \chictwo\ are summarized
and presented as inclusive $B$
branching fractions in Table~\ref{tab:brsum}.

\begin{table}[!htb]
\caption{Summary of inclusive $B$ branching ratios measured
with respect to $B\rightarrow\jpsi X$.
The total systematic error in the \epem\ and \mumu\ final states is
the sum in quadrature of the final-state-specific values and the systematic error
common to both. The results are combined and multiplied by the 
PDG value for $\BR(B\rightarrow\jpsi X)$ to obtain inclusive
branching fractions (\%).} 
\label{tab:brsum}
\vskip 0.1in

\begin{center}
\begin{tabular}{|l|ccc|ccc|c|ccc|} \hline
   & \multicolumn{7}{|c|}{Branching ratio relative to \jpsi} &
     \multicolumn{3}{|c|}{Branching fraction (\%)} \\ \hline
   & \multicolumn{3}{|c|}{\epem} &  \multicolumn{3}{|c|}{\mumu}
   & Common &  \multicolumn{3}{|c|}{Combined} \\ 
   &    & Stat & Sys &      & Stat & Sys & Sys (\%) &     & Stat & Sys \\
   \hline \hline
$\chicone\rightarrow\gamma\jpsi$ & 0.28 & 0.06 & 0.03 & 0.38 & 0.05 & 0.02 &
\phantom{0}7.3 & 0.39 & 0.04 & 0.04 \\ \hline
\chictwo\, 90\% CL & 0.16 & & &0.28 & & & \phantom{0}8.1 &0.24 & & \\ \hline
$\psitwos\rightarrow\ell^+\ell^-$ & 0.22 & 0.05 & 0.01 & 0.23 & 0.04 & 0.01 &
13.6 & 0.26 & 0.03 & 0.04 \\ 
$\psitwos\rightarrow\pipi\jpsi$ & 0.18 & 0.06 & 0.01 & 0.22 & 0.03 & 0.00
&\phantom{0}9.9 & 0.24 & 0.03 & 0.03 \\ 
Combined \psitwos & & & & & & & & 0.25 & 0.02 & 0.02 \\ \hline
\end{tabular}
\end{center}
\end{table}

\section{Summary}
\label{sec:summary}

The charmonium states \jpsi, \psitwos\ and \chic\ have been
reconstructed in $B$ meson decays.  
Preliminary measurements of the ratio of branching fractions 
$\BR(B\rightarrow \psitwos X)/\BR(B \rightarrow \jpsi X)$ and 
$\BR(B\rightarrow \chicone X)/\BR(B \rightarrow \jpsi X)$ have been
presented and a
limit set on the ratio for \chictwo.  Using the Particle Data Group value 
for the
inclusive branching fraction for \jpsi, these values have been converted into
inclusive $B$ branching fractions.

A direct value for the \jpsi\ branching fraction will be available upon
completion of particle identification studies. 
Several of these results
are statistically limited; more precise values will be available based on 
data that are currently being recorded.  The new data will also
permit measurements of the $p^*$ and helicity distributions for
these charmonium mesons.

\section{Acknowledgments}
\label{sec:Acknowledgments}

We are grateful for the contributions of our \pep2\ colleagues in
achieving the excellent luminosity and machine conditions
that have made this work possible.
We acknowledge support from the
Natural Sciences and Engineering Research Council (Canada),
Institute of High Energy Physics (China),
Commissariat \`a l'Energie Atomique and
Institut National de Physique Nucl\'eaire et de Physique des Particules
(France),
Bundesministerium f\"ur Bildung und Forschung
(Germany),
Istituto Nazionale di Fisica Nucleare (Italy),
The Research Council of Norway,
Ministry of Science and Technology of the Russian Federation,
Particle Physics and Astronomy Research Council (United Kingdom), the
Department of Energy (US),
and the National Science Foundation (US). In addition, individual support 
has been received from the Swiss 
National Foundation, the A. P. Sloan Foundation, the Research Corporation,
and the Alexander von Humboldt Foundation.
The visiting groups wish to thank 
SLAC for the support and kind hospitality
extended to them.

\end{document}